# Virtualizing Lifemapper Software Infrastructure for Biodiversity Expedition

Nadya Williams
University of California, San Diego
9500 Gilman Dr., MC 0444
La Jolla, CA 92093, USA
+1 858-534-1820
nadya@sdsc.edu

Aimee Stewart
KU Biodiversity Institute
1345 Jayhawk Blvd
Lawrence, KS 66045, USA
+1 785-864-2233
aimee.stewart@ku.edu

Phil Papadopoulos
University of California, San Diego
9500 Gilman Dr., MC 0505
La Jolla, CA 92093, USA
+1 858-822-3628
phil@sdsc.edu

## ABSTRACT
One of the activities of the Pacific Rim Applications and Grid Middleware Assembly (PRAGMA) is fostering Virtual Biodiversity Expeditions (VBEs) by bringing domain scientists and cyber infrastructure specialists together as a team. Over the past few years PRAGMA members have been collaborating on virtualizing the Lifemapper software. Virtualization and cloud computing have introduced great flexibility and efficiency into IT projects. Virtualization provides application scalability, maximizes resources utilization, and creates a more efficient, agile, and automated infrastructure. However, there are downsides to the complexity inherent in these environments, including the need for special techniques to deploy cluster hosts, dependence on virtual environments, and challenging application installation, management, and configuration. In this paper, we report on progress of the Lifemapper virtualization framework focused on a reproducible and highly configurable infrastructure capable of fast deployment.

Lifemapper is a complex biological software infrastructure developed by the Biodiversity Institute at The University of Kansas that creates and maintains an archive of species distribution maps calculated from public specimen data and a suite of data and tools for biodiversity researchers that calculate single and multi-species distribution predictions and macro-ecological analyses. Our goal is to create a viable virtualization solution that can be easily adopted and reused by scientists at multiple institutions and projects. This solution 1) allows fast deployment of ready-made cluster images; 2) reproduces the complete Lifemapper processing pipeline on demand at multiple sites and in different hosting environments; and 3) enables scientists to perform Lifemapper-facilitated data processing on restricted-use data, very large datasets, or other unique data.

A key contribution of this work is describing the practical experience in taking a complex, clustered, domain-specific, data analysis and simulation system and making it available to operate on a variety of system configurations. Uses of this portability range from whole cluster replication to teaching and experimentation on a single laptop. System virtualization is used to practically define and make portable the full application stack, including all of its complex set of supporting software.

## Categories and Subject Descriptors
D.2.11 [**Software Engineering**]: Software Architectures – *domain-specific architectures, patterns*.

## General Terms
Design

## 1. INTRODUCTION
Data on species distributions is a core resource for biodiversity studies and using it effectively requires innovation in the way it is managed for research analyses. The Lifemapper Project [1] has innovated processing and analysis of species information resulting in a modular platform capable of modeling of individual species distributions as well as models of patterns of biological diversity. Analyzing continental and global scale properties and patterns of the biological diversity of natural communities of plant and animal species involves the creation of large data sets derived from models of species distribution which are in turn computed from historical species occurrence data and observations on species. These data sets are ideal for research in macroecology, the science of analyzing biological diversity on continental and global scales and has proven useful for revealing key biological and geographical features of the distribution and diversity of species.

Currently only Lifemapper developers have a good understanding of the software installation, initial data population, communications setup and usage, but both institutions and scientific researchers have requested individualized Lifemapper instances to communicate with a local or remote compute resource containing restricted or specialized data for their own archive. Virtualizing the software components greatly aided and simplified the steps for creating a targeted Lifemapper instance.

## 2. LIFEMAPPER – MAIN COMPONENTS
Lifemapper is a complex biological software infrastructure consisting of three independent components that communicate with each other to process biological data: (1) LmServer for data management and communications; (2) LmCompute for calculations; and (3) various client applications including a plugin to the open source GIS application QGIS [10]. Lifemapper data is made up of two primary types of data - user data and the Lifemapper Archive. To create the Lifemapper Archive, the data pipeline starts with species occurrence data from a data aggregator (e.g. GBIF, iDigBio, BISON) and computes Species Distribution Models (LmSDM) based on accepted taxonomic names as defined by the provider. This occurrence data is combined with observed or modeled climate layers and processed to produce the LmSDM portion of the Lifemapper Archive, freely available for query and download. LmCompute and LmServer share these operations. LmCompute requests jobs from LmServer, executes them, then posts the results back to LmServer where data are written to storage and metadata to the PostgreSQL database. Two applications on LmCompute underlie SDM calculations: openModeller [5][8][9] and MaxEnt [4][7]. From LmServer, a user can access original and computed species data and metadata through the Lifemapper website or with one a client, such as the QGIS plugin. Figure 1 illustrates the separation between the two primary components, LmServer, on which the Data Pipeline continuously updates species data, assembles job packages for computation and stores results in the Lifemapper Archive, and LmCompute, on which processes are executed.

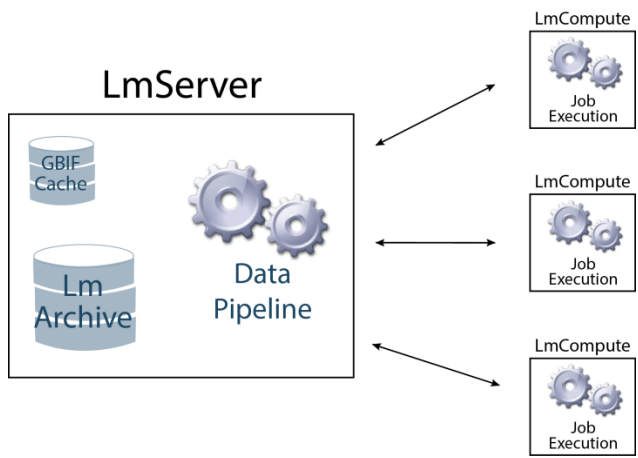

**Figure 1. Lifemapper architecture.**

A primary goal is to make Lifemapper components portable, easily deployable and scalable. We achieved this via incremental virtualization of the Lifemapper components. Bringing "big data" or data with usage restrictions into a virtualized environment further complicates implementation. One strategy for handling big or restricted data is to bring computations to the data: employing virtual servers and clusters to run applications in the location of the data rather than bringing the data to an application.

## 3. LIFEMAPPER VIRTUALIZATION
### 3.1 Software packaging as a Rocks roll
The original Lifemapper install was on Ubuntu-based hosts at the Biodiversity Institute in the Kansas University. Like most legacy system that are fairly complex it requires a large set of dependencies to be installed and configured, and all compilation and configuration was done using homegrown scripts and developers knowledge.

As a first step, we chose to move the Lifemapper installations to Rocks clusters [6] because the Rocks toolkit allows us to provision a physical (or virtual) cluster in an efficient, programmable, configurable way in a very short time. The resulting robust cluster is in a known state and has a known modular software stack and with automatic configuration of database, web server and job scheduler components needed for various Lifemapper components.

We reviewed the installation at Kansas University's Biodiversity Institute and created a build process enabling a fast turnaround from software update to server availability. For this, we completed full refactoring of the Lifemapper software stack and automated the software build and install process via Rocks rolls. Rocks roll [2] is a powerful method to add software packages to the cluster; a roll contains software packages as RPMs, possible extensions to cluster command line structure, and a set of instructions to automatically configure the software based on the cluster node functionality or role and on its physical layout. Packaging an application as a roll greatly facilitates its installation, configuration and update and enables a convenient sharing of the resulting software roll with other users. Creating the application roll enables the application portability and deployment in reproducible and reliable way and makes application integration into the cluster seamless and automatic. For Lifemapper, we chose to create separate rolls for the LmCompute and LmServer components.

### 3.2 LmCompute virtualization
The first logical component for virtualization was LmCompute. For PRAGMA 25 our goal was to separate the components and deploy LmCompute as a virtual cluster at SDSC.

We created a *lifemapper* roll for LmCompute and deployed several instances of it as separate clusters to handle jobs with unique software or data requirements. The roll installs all the required software and prerequisites for this component, and configures the cluster to use a specific LmServer host. After the roll is installed, the cluster is ready to run Lifemapper jobs. The cluster frontend pulls the jobs from LmServer and dispatches them to the compute nodes via a job submitter script.

Roll building and installing is mostly automated so recreating a roll with a software update and applying it to an existing installation is now a much more efficient and trouble free process. The roll packaging system reduces the cost of installing, configuring and replicating the LmCompute component. Figure 2 illustrates how the software integration and installation complexity is now encapsulated in the development-production pipeline:

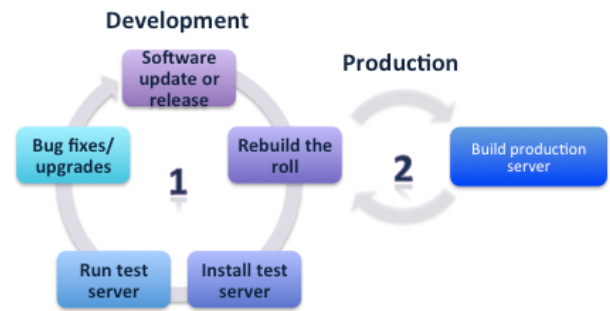

**Figure 2. Rocks rolls reduce the time between software updates and server availability.**

The roll helped to drastically reduce the time spent on software build and configuration and automated nearly all hands-on tasks eliminating the burden of integration of hardware and software for the compute component. Building and testing a new compute resource became trivial and there is little difference between building virtual or physical resources because Rocks treats virtual hardware as another type of physical hardware. LmCompute roll source is available on PRAGMA's github (github.com/pragmagrid/lifemapper). Using the roll we can now create a custom, portable and easily deployable cluster. During PRAGMA 25 we demonstrated our streamlined and automated build process, deployed a virtual cluster at SDSC that was running LmCompute component and processing Lifemapper jobs that were created on the original, non-refactored, LmServer at The University of Kansas. In addition, we extended the code to produce process-specific provenance logging for Lifemapper jobs using Indiana University's KARMA provenance collection tool.

### 3.3 LmServer Virtualization

As the Lifemapper project considers more efficient data storage and query, it will need to experiment with different physical disks, dataset organizations and layouts, and file formats. This will require a few instances of a portable and reproducible LmServer to test under various conditions. While essential for researching the best strategies for the data store, a portable LmServer also addresses two other specific needs. First, it allows colleagues in the PRAGMA VBE at the University of Florida to use recently acquired satellite data licensed only to physical storage at Florida. Florida researchers requested a local installation of Lifemapper to compute high quality species models using this restricted satellite data. A portable LmServer would complete the Lifemapper system, and allow data assembly, cataloging, storage, and computation to operate independently at a location other than its origin in Kansas. Second, other data aggregators would benefit from having a self-contained Lifemapper system that could operate solely on their data. These institutions desire a local copy of the LmServer to manage data storage, retrieval, data and analysis web services, and website, to create their own institutional SDM or RAD archive focused solely on their data or region. These LmServer installations can be self-contained (installed locally with LmCompute on an institutional cluster) or can be distributed and utilize any available Lifemapper LmCompute installations.

These specific needs are solved by additional Lifemapper virtualization. For PRAGMA 26, we decoupled the web server and database server from the KU-specific implementation and created a *lifemapper-server* roll (source is available at github.com/pragmagrid/lifemapper-server) enabling a build process for a configurable and easily deployable virtual server.

The server software has many dependencies to be satisfied both during roll build time and during installation. In our lifemapper-server roll we created an end-to-end build, install and configure process that automates the entire lifecycle of application management and enables: (1) fast software updates or rollback; (2) simple packaging and reliable robust deployment; (3) VM provisioning where building a virtual host is no different than building a physical host; (4) a hardened installation process and full integration with the underlying cluster via customizable configuration files.

For PRAGMA 27, we automated Lifemapper server data and metadata seeding for a test dataset, then initiated data computation on the LmCompute, and confirmed that LmServer provided jobs to LmCompute and data to end users.

We now have an installation where both the data management and communication functions of LmServer can reside on the same or different hosts using the same lifemapper-server roll. This provides portability, easy deployment, and the flexibility to make custom installations depending on the specific site needs.

### 3.4 Using Different Virtualization Technologies

The generalized approach for running scientific applications in PRAGMA is to create virtual machines (VMs) with all software to run the applications and to start VM instances at new sites. This approach is intended to leverage the infrastructure and application virtualization and to reduce the complexity and labor-intensive application management and host instantiation. Many PRAGMA sites can participate as resource providers and deploy VMs prepared elsewhere on their infrastructure. One of our goals is to support the portable Lifemapper server using different virtualization technologies. The first virtualization solution we chose is Kernel-based Virtual Machine (KVM [3]), ideal for Rocks-based clusters. Our second virtualization solution involves creating a VM with VirtualBox [11] on a laptop. Laptop installations are the first step towards allowing scientific expeditions to use Lifemapper in the field.

*3.4.1 KVM*

KVM is a full virtualization solution for Linux on x86 platform and is ideal for clusters. Rocks cluster management provides tools via the KVM roll for virtual cluster management and deployment. The Rocks KVM roll allows users to create and deploy virtual hosts and virtual clusters reliably and efficiently and then add previously created software rolls as needed. We can build multiple virtual clusters with identical or different setups depending on the desired cluster configuration. Running virtual clusters in KVM for the Virtual Biodiversity Expedition resulted in multiple advantages: (1) larger instance sizes that are limited only be the hosting hardware specification; (2) long lasting instances used by multiple external clients; (3) dynamic input data; (4) multiple virtual clusters; (5) dynamically grow clusters based on computational needs.

*3.4.2 VirtualBox*

VirtualBox, a free and Open Source software, is a powerful x86 and AMD64/Intel64 virtualization product that runs on Linux, Windows, OSX and is ideal for laptops. Running virtual images on a laptop has some drawbacks. For some users, networking setup may not be trivial. In addition, laptop hardware often limits the storage and memory available to virtual images. However, benefits offset these limitations. The VBEs need special purpose or short-lived instances as well as unique input data. VirtualBox instances are also well suited for teaching or situations there is no network connection. The instantiation of virtual cluster ready-made images can be accomplished in very few steps.

*3.4.3 Virtualization scenarios*

While multiple virtualization technologies provided us with different virtual clusters deployment options, the next challenge was to provide a consistent implementation that poses very few challenges for the users; easy to adopt and install. Our solution was to create Rocks Virtual Clusters for both KVM and VirtualBox, allowing developers to develop in the same environment as production systems are deployed. While the provisioning of the cluster is different in KVM and VirtualBox, the image building can have the same flow. In addition, virtualized images allow us to leverage the PRAGMA developed tools *pragma_boot* and *cloud scheduler* for automated cluster startup, overlay networks or Software Defined Networks (SDNs). These tools set up data networks and create virtual machines and clusters with the capability to be deployed in EC2.

For PRAGMA 27 we created KVM and VirtualBox clusters and deployed them in the following 3 scenarios:

1. Two KVM virtual clusters, LmCompute and LmServer at SDSC on physical hardware running Rocks 6.1.1.
2. A single VirtualBox cluster with both LmServer and LmCompute components installed on a laptop.
3. Two VirtualBox clusters, LmServer and LmCompute at SDSC

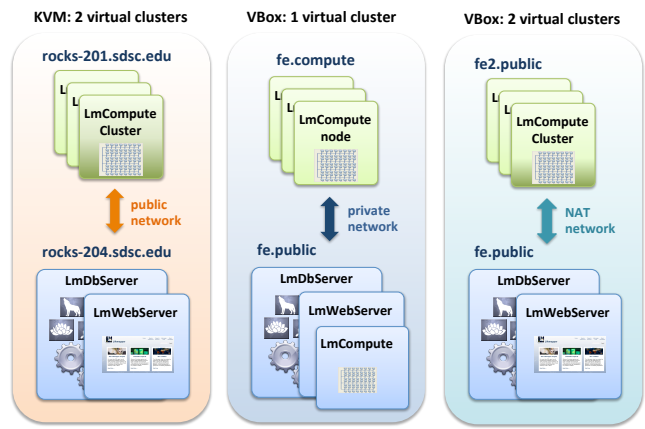

Figure 3. Lifemapper Virtual Clusters Scenarios for VBE.

We use Rocks rolls for all cluster software encapsulating difficult application installation and configuration. With the rolls, we know our system real time status: installed software, versions, and configurations. The resulting robust system can be reliably built and rebuilt.

By creating virtual machines and virtual clusters using Rocks in both KVM and VirtualBox virtualization environments, we virtualized the infrastructure improving hardware utilization and scalability. The flexibility and efficiency of virtual servers allowed easier testing and experimenting because the virtual images can be easily replicated, exported and moved to other sites, facilitating deployment on shared PRAGMA cloud resources and allowing for deployment of multiple LmCompute clusters across the multiple sites as needed.

## 4. Distributed computing and Geographically Restricted Data Resources

The sharing of software, data and computational capabilities across international networks in a trusted environment provides a great opportunity for biodiversity and other researchers. We must adapt technologies to be responsive to international agreements on data restrictions, use and distribution. The integration of data resources with virtual clusters becomes a vital component of VBE and PRAGMA is committed to address the technical issues of data and application sharing among collaborators and hosting sites through virtualization.

For PRAGMA 28, we addressed both geographically restricted data resources, and geographically distributed software components. Commercial satellite imagery data from Kinabalu (Borneo, Malaysia) was geographically restricted to the University of Florida (UF). The PRAGMA Testbed sites were used for instantiating distributed Lifemapper software components. SDSC created Lifemapper virtual images for all components then migrated the LmServer virtual cluster to another site at SDSC and the LmCompute virtual cluster to the UF site. Both new clusters were re-instantiated at their respective new sites. The restricted data was installed on a host at UF that is not a part of the PRAGMA Testbed then connected to the LmCompute virtual cluster using Vine private network.

Figure 4 illustrates the infrastructure used to distribute Lifemapper components and connect them to restricted data:

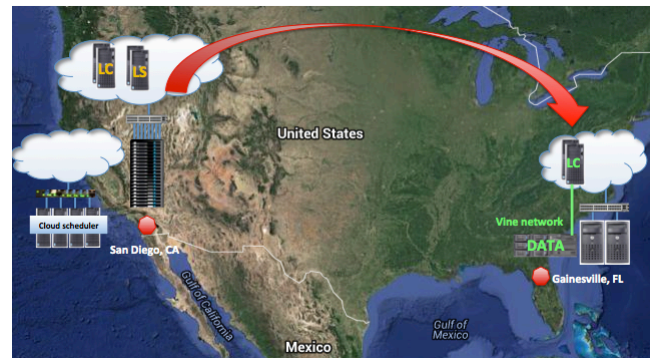

Figure 4. Infrastructure for Lifemapper virtualization.

We deployed LmCompute at the same location as the restricted input data, while the PRAGMA network setup simplified authentication and data permissions. The resulting Lifemapper workflow is illustrated in Figure 5:

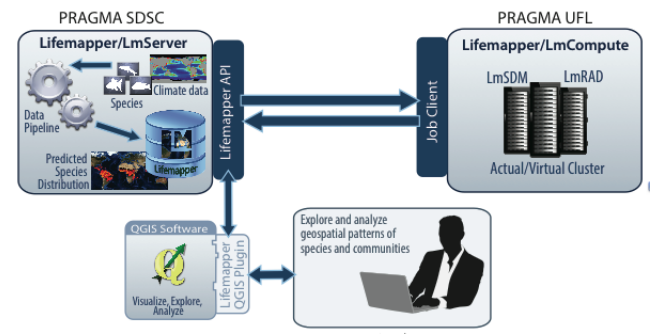

Figure 5. Lifemapper on PRAGMA Testbed.

## 5. SUMMARY

The PRAGMA Virtual Biodiversity Expedition (VBE) is designed to use advanced technology to address pressing biological questions about species distributions and adaptations in extreme environments. This paper described a programmatic approach to creating and deploying virtual images on the PRAGMA Testbed.

Automating development cycles via Lifemapper rolls we are able to improve the quality of the applications and easily install Lifemapper on physical or virtual clusters on demand. We can use the same ISO image to install the Lifemapper components on physical or virtual clusters and reduce the time cost from hours and days needed to build the software by hand to 15 minutes of roll installation. We automated time-intensive and repetitive tasks, reducing the risks associated with software releases and updates and facilitating faster development and scale up. The benefits of creating virtual cluster images result from using a well defined build process from development to production deployment, seamlessly integrating software and hardware.

The virtual machines and clusters can be used for real time experiments as well as training mechanisms. The "make once, eat all week" approach allows us to create a complete system as an end-to-end solution while greatly reducing the cost of installing, configuring and replicating.

We envision following this experiment with more general use-cases deploying Lifemapper and other science applications in dynamic workflows. This approach can serve as a blueprint and used for other scientific applications and Virtual Biodiversity Expeditions and provide this new installation capability to other users.

All the source code for the rolls and the Lifemapper components is publically available on the github to PRAGMA and non-PRAGMA users alike.

## 5.1 Future work

Our goal in this work is to create a viable virtualization solution that can be easily adopted and reused by scientists at other institutions and projects. Several next steps are already underway:

- Continue modularizing Lifemapper code to accommodate alternate input data by simplifying data initialization and population. We must formalize the requirements for fully described data allowing easy use of different input datasets (iDigBio, GBIF, BISON, individual researcher, etc.) and switching among them. Extend the pipeline to enable dynamic multi-species pattern analyses of a Lifemapper instance populated with data of Mount-Kinabalu, Malaysia. New Lifemapper modules enabling batch processing, editing pipeline workflows, spatial queries and archive subsets for dynamic macroecological analysis will facilitate more complete biodiversity analyses of the region. The BISON and iDigBio user communities will be able to create Lifemapper installations on their own once this modularization is complete.
- Create an infrastructure bridging Indonesia and other PRAGMA sites with a dedicated LmServer for our Indonesian colleagues.
- Finalize the networking between LmServer and LmCompute components on a single VC on a laptop. This will facilitate the use of Lifemapper in the field or offline with newly collected field data.
- Incorporate different networking scenarios in the Lifemapper virtual infrastructure facilitated by advances in overlay networks in the PRAGMA Experimental Network Testbed. This can enable a deployment scenario when a single LmCompute cluster is distributed across multiple sites.

## 6. ACKNOWLEDGMENTS

This work is funded in part by National Science Foundation (PRAGMA grant number 1234953, Lifemapper grant numbers BIO/ABI 1356732, BIO/ABI 1458422, Rocks grant numbers OCI-1032778 and OCI-0721623, iDigBio grant number EF-1115210) and USGS (Lifemapper grant number BISON G14AC00285).